\documentclass[amsmath, amssymb, aps, floatfix, nofootinbib, preprintnumbers, showpacs,	twocolumn] {revtex4}

\usepackage{dcolumn}								
\usepackage[dvips]{graphicx}							

\begin{document}

\title{ Cosmology with Hu-Sawicki gravity in Palatini Formalism}

\author{B. Santos${^1}$\footnote{thoven@on.br}}

\author{M. Campista$^1$\footnote{campista@on.br}}

\author{J. Santos$^2$\footnote{janilo@dfte.ufrn.br}}

\author{J. S. Alcaniz${^1}$\footnote{alcaniz@on.br}}

\affiliation{$^1$Observat\'orio Nacional, 20921-400 Rio de Janeiro - RJ, Brasil}

\affiliation{$^2$Departamento de F\'{\i}sica, Universidade Federal do Rio Grande do Norte,  59072-970 Natal - RN, Brasil}

\date{\today}

\begin{abstract}
Cosmological models based on $f(R)$-gravity may exhibit a natural acceleration mechanism without introducing a dark energy component. In this paper, we investigate cosmological consequences of the so-called Hu-Sawicki $f(R)$ gravity in the Palatini formalism. We derive theoretical constraints on the model parameters and perform a statistical analysis to determine the parametric space allowed by current observational data. We find that this class of models is indistinguishable from the standard $\Lambda$CDM model at the background level. Differently, from previous results in the metric approach,  we show that these scenarios are able to produce the sequence of radiation-dominated, matter-dominated, and accelerating periods without need of dark energy.
\end{abstract}


\maketitle

\section{Introduction}

General Relativity (GR) is a very well-tested and
established theory of gravity. However, all the successful tests performed so
far do not account for the ultra-large scales corresponding to the low curvature
characteristics of the Hubble radius today. Thus, in principle it is conceivable
that the recent discovery of the late-time cosmic acceleration~\cite{obs_data}
could be explained by modifying Einstein's GR in the far infrared regime.
Although the diversity of approaches is the hallmark in this field (see, e.g.,
Refs.~\cite{Modified_Gravity}), the simplest possible theory results by adding
terms proportional to powers of the Ricci scalar $R$ to the Einstein-Hilbert
Lagrangian, the so-called  $f(R)$ gravity (see Refs.~\cite{Sotiriou} for
recent reviews).

Differently from the standard general relativistic scenario, $f(R)$ cosmology can
naturally drive an accelerating cosmic expansion without introducing dark energy.
However, the freedom in the choice of different functional forms of $f(R)$ gives rise to the problem of
how to constrain the many possible $f(R)$ gravity theories. In this regard, much
efforts have been developed so far, mainly from the theoretical
viewpoint~\cite{theoretical_issues}. General principles such as the so-called
energy conditions~\cite{energy_conditions}, nonlocal causal
structure~\cite{causal_structure}, and issues such as loop quantum
cosmology~\cite{quantum_gravity}, have also been taken into account in order to
clarify its subtleties. More recently, observational constraints from several
cosmological data sets have also been explored for testing the viability of these
theories~\cite{cosmological_tests}.

An important aspect that is worth emphasizing concerns the two different
variational approaches that may be followed when one works with $f(R)$ gravity,
namely, the metric and the Palatini formalisms. In the metric formalism the
connections are defined \textit{a priori} as the Christoffel symbols of the
metric and the variation of the action is taken with respect to the metric,
whereas in the Palatini variational approach the metric and the affine
connections are treated as independent fields and the variation is taken with
respect to both (for a review on $f(R)$ theories in the Palatini approach
see~\cite{Olmo1}). The result is that in the Palatini approach the connections
depend on the metric and also on the particular $f(R)$, while in the metric
formalism the connections depend only on the metric. We have then that the same
$f(R)$ leads to different spacetime structures.

These differences also extend to the observational aspects. For instance, we
note that some cosmological models based on a power-law functional form in the
metric formulation fail in reproducing the standard matter-dominated era
followed by an acceleration phase~\cite{Amendola} (see, however, \cite{cap}),
whereas in the Palatini approach, some analysis have shown that such theories admit the three post
inflationary phases of the standard cosmological model~\cite{Tavakol}.
Nevertheless, we do not have yet a clear comprehension of the properties of the
Palatini formulation of $f(R)$ gravity in other scenarios, and issues such as
the Newtonian limit~\cite{Meng-Wang} and the Cauchy problem~\cite{Lanahan} are
still contentious (see, however, Refs.~\cite{Cauchy-Problem}). Another issue
with the Palatini $f(R)$ formulation concerns surface singularities of static
spherically symmetric objects with polytropic equation of
state~\cite{Barausse, Olmo1}. This problem has been reexamined in
Refs.~\cite{Kainulainen}, where the authors showed that the singularities may
have more to do with peculiarities of the polytropic equation of state used
(e.g. its natural regime of validity) and the $f(R)$ model chosen.

Although being mathematically  simpler than the metric formulation, the Palatini approach has received little attention from the point of view of cosmological tests. Following early works in this direction~\cite{Tavakol,janilo,expogravity}, in this paper we explore the cosmological consequences of a class of modified $f(R)$ gravity, recently proposed by Hu \& Sawicki~\cite{HS}, in the Palatini formulation. Among a number of $f(R)$ models discussed in the literature,  this is designed to posses a chameleon mechanism that allows to evade solar system constraints and the cosmological scenario that arises from the metric formalism of this model has been shown to satisfy the conditions needed to produce a cosmologically viable expansion history. In order to test the observational viability of this class of models in the Palatini approach, we use different kinds of observational data, namely: type Ia supernovae (SNe Ia) observations from Union2.1 sample~\cite{union2.1},  estimates of the expansion rate at $ z \neq 0 $, as discussed in Ref.~\cite{newh}, current measurements of the product of the CMB acoustic scale $\ell_A$ and the comoving sound horizon at photon decoupling (CMB/BAO ratio)~\cite{cmbbao} and measurements of the gas mass fraction in galaxy clusters~\cite{fgas}. We show that for a subsample of model parameters, the so-called Hu-Sawicki scenarios in the Palatini approach are indistinguishable from the standard $\Lambda$CDM model, being able to produce the sequence of radiation-dominated, matter-dominated, and accelerating periods without need of dark energy.

\section{The Palatini approach}

The modified Einstein-Hilbert action that defines an $f(R)$ gravity is given by
\begin{equation}
 S = \frac{1}{2\kappa^2}\int d^4x\sqrt{-g}f(R) + S_m \;,
\end{equation}
where $ \kappa^2 = 8 \pi G $, $g$ is the determinant of the metric tensor and $S_m$ is the standard action for the matter fields.

In the Palatini variational approach the fundamental idea is to regard the
torsion-free connection $\Gamma^{\rho}_{\mu\nu}$, as well as the metric
$g_{\mu\nu}$ entering the action above, as two independent fields.
The field equations in this approach are
\begin{subequations}
\label{eqs}
\begin{equation}
	\label{eq:field_eqs}
	f_R R_{(\mu\nu)} - \frac{f}{2}g_{\mu\nu} = \kappa^2T_{\mu\nu}
\end{equation}
and
\begin{equation}
         \label{eq:field_eqsb}
         	 \tilde{\nabla}_\beta\left( \sqrt{-g}\,f_R g^{\mu\nu}\right) = 0 \;,
\end{equation}
\end{subequations}
where $ f_R = df/dR $ and $\tilde{\nabla}$ represents covariant derivatives.
From the above equations, we obtain the connections
\begin{equation}
	\label{connections}
	\Gamma_{\mu\nu}^{\rho} = \left\{^{\rho}_{\mu\nu}\right\}
				+ \frac{1}{2f_R}\left( \delta^{\rho}_{\mu}\partial_{\nu}
				+ \delta^{\rho}_{\nu}\partial_{\mu}
				- g_{\mu\nu}g^{\rho\sigma}\partial_{\sigma} \right)f_R \,,
\end{equation}
where $\left\{^{\rho}_{\mu\nu}\right\}$ are the Christoffel symbols of the
metric. As mentioned earlier, the connections, and so the gravitational fields,
are described not only by the metric $g_{\mu\nu}$ but also by the proposed
$f(R)$ theory. As a cosmological model we consider a homogeneous isotropic
Universe described by the Friedmann-Lema\^{i}tre-Robertson-Walker (FLRW) flat
geometry $ g_{\mu\nu} = diag(-1,a^2,a^2,a^2) $, where $a(t)$ is the cosmological
scale factor, and as the source of curvature a perfect-fluid with energy density
$\rho$ and pressure $p$ whose  energy-momentum tensor is given by
$ T^{\mu}_{\nu} = diag(-\rho,p,p,p) $.

The generalized Friedmann equation, obtained from (\ref{eqs}) and
(\ref{connections}), can be written in terms of the redshift $z$ as 
\begin{equation}
	\label{eqf}
	\frac{H^2}{H_0^2} = \frac{
					3\Omega_{m0} (1+z)^3 + 6\Omega_{r0}(1+z)^4 + f(R)/H_0^2
				}{
					6f_R \xi^2
				} \,,
\end{equation}
where
\begin{equation}
	\xi = 1 + \frac{9}{2} \frac{ f_{RR} }{ f_R }
			\frac{ H_0^2\Omega_{m0}(1+z)^3 }{ Rf_{RR} - f_R } \,,
\end{equation}
and $\Omega_{m0}$ and $\Omega_{r0}$ stand for the present-day values of the matter and radiation density parameters. The trace of (\ref{eq:field_eqs}) results in
\begin{equation}
	\label{eq:trace}
	Rf_R - 2f = -3H_0^2\Omega_{m0}(1+z)^3 \,,
\end{equation}
which we will consider in order to restrict the parameters of the given $f(R)$
theory.
\\

\begin{figure*}[t]
	\includegraphics[width= 7.5cm, height=6.0cm] {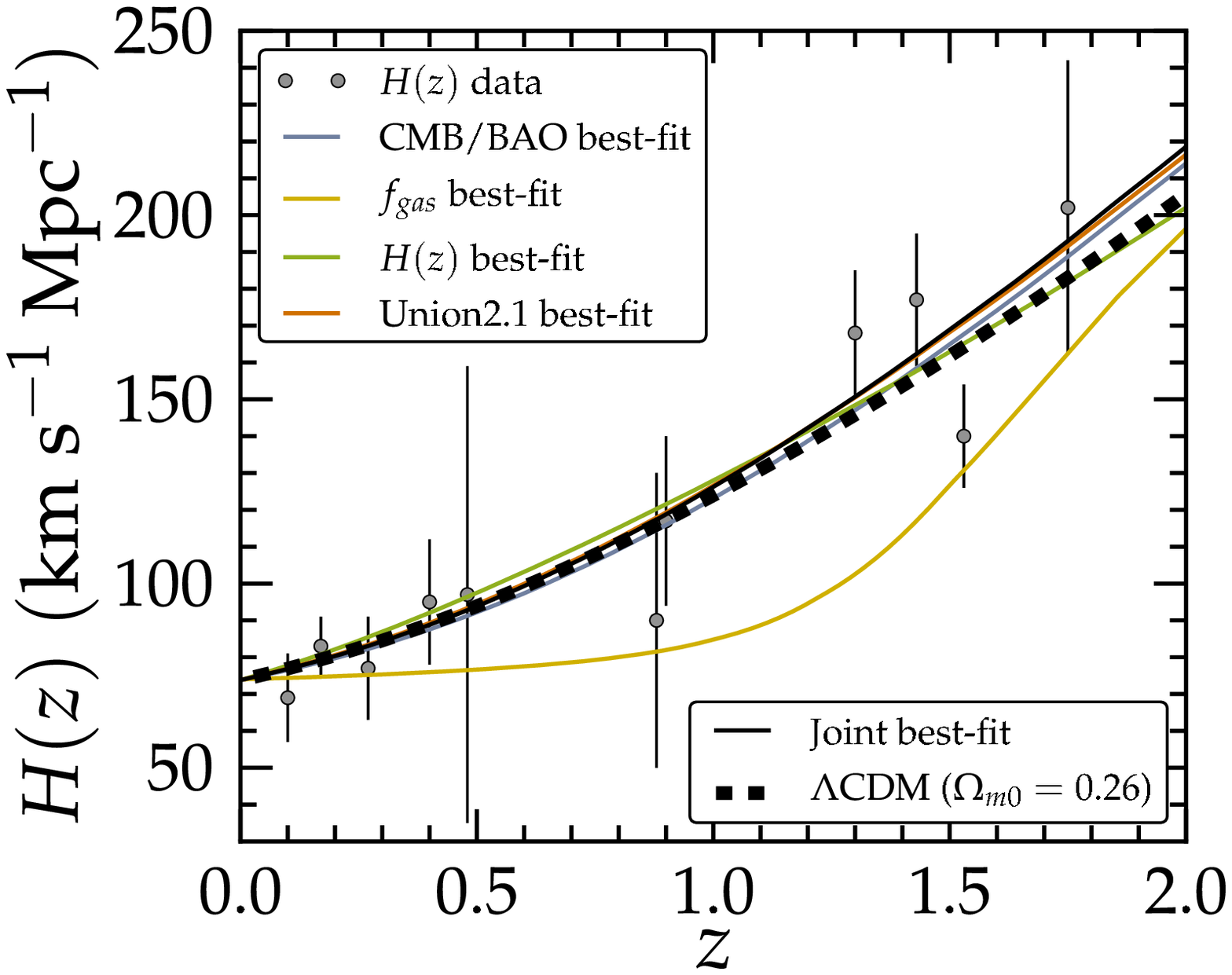}
\hspace{0.4cm}
	\includegraphics[width= 7.5cm, height=6.0cm]{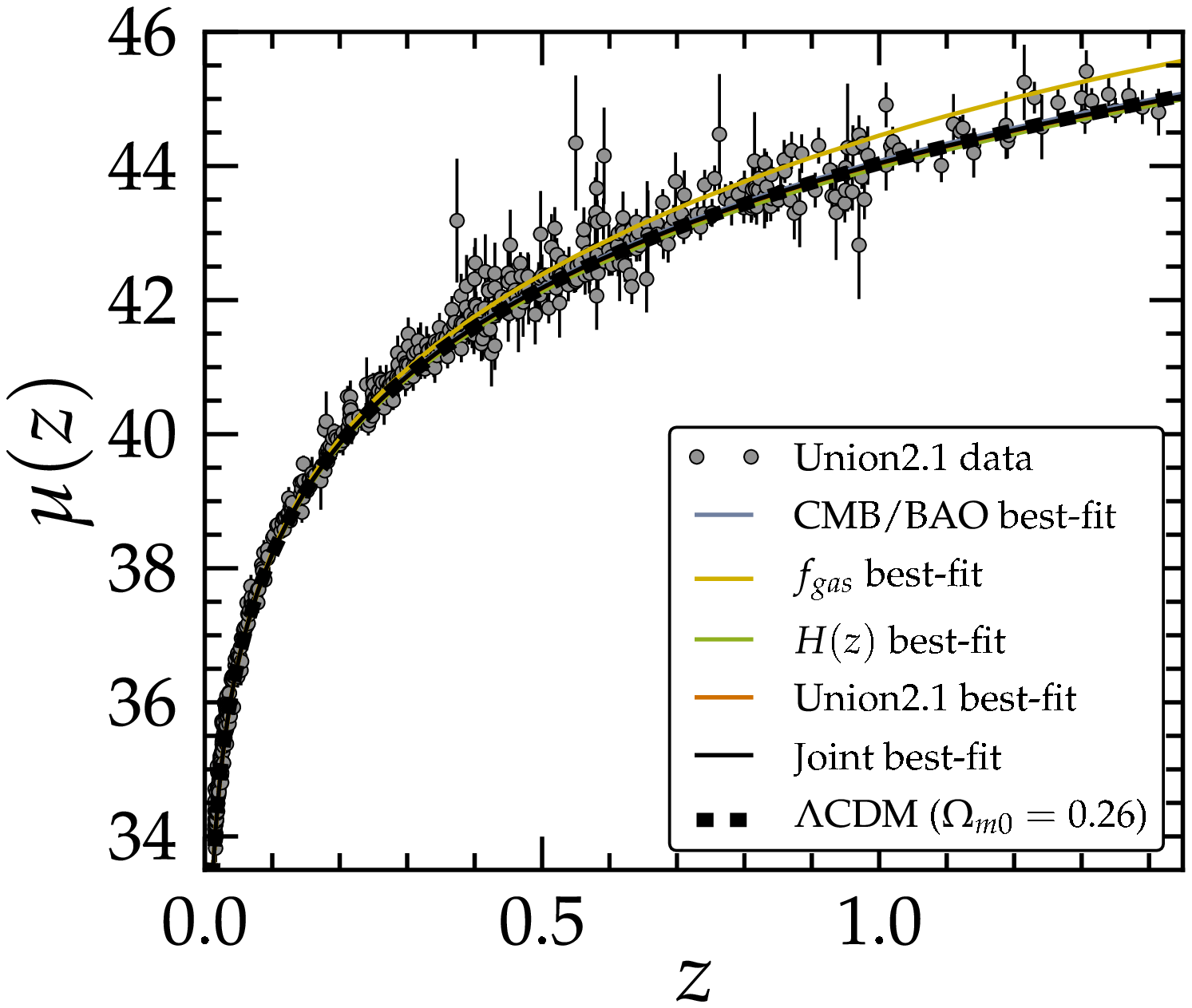}
	\caption{\label{fig:data_h-sn}
		(\textit{left}) The predicted Hubble evolution $H(z)$ as a
		function of the redshift for the Hu-Sawicki model in the
		Palatini formalism (Eqs. \ref{eqf}-\ref{Hu-Sawicki}). The curves
		correspond to the best-fit values of $\Omega_{m0}$ and $c_2$
		discussed in the text. For the sake of comparison, the standard
		$\Lambda$CDM model prediction is also shown. The data points are
		the measurements of the $H(z)$ given in Ref.~\cite{newh}.
		(\textit{right})~Hubble diagram for 580 SNe Ia from the Union2.1
		sample~\cite{union2.1}. The curves correspond to the best-fit
		values of $\Omega_{m0}$ and $c_2$ displayed in
		Table~\ref{tab:results}.
	}
\end{figure*}

\subsection{The Hu-Sawicki gravity model}

In this paper we are interested in the $f(R)$ modified gravity model proposed by Hu and Sawicki~\cite{HS}
\begin{equation}
	\label{Hu-Sawicki}
	f(R) = R -m^2 \frac{ c_1 \left(\frac{R}{m^2}\right)^n }{ c_2 \left( \frac{R}{m^2} \right)^n + 1 } \,.
\end{equation}
For the case $n = 1$, the above equation can be rewritten as~\cite{hu_cluster}
$$
f(R) = R - \frac{2\Lambda}{ \frac{\mu^2}{R} + 1}\;,
$$
where $ \Lambda = m^2 c_1 / 2c_2 $ and $ \mu^2 = m^2 / c_2 $ are dimensionless parameters and $ m^2 = H_0^2 \Omega_{m0}$. Note that, in the regime $ R >> \mu^2 $, the model above is practically indistinguishable from the $\Lambda$CDM scenario\footnote{As shown in Ref.~\cite{hu_cluster}, this is equivalent to have $ |f_{R0}| \sim 1 $, where the subscript 0 denotes present-day quantities. See also Table I for some estimates of this quantity.}. We have verified that the parameter $n$ is completely unconstrained by current data. Thus, without loss of generality and following  Ref.~\cite{hu_cluster}, we focus our analyses on the case $n = 1$.

The model above is intended to explain the current acceleration of the universe without considering a cosmological constant term. Cosmological and astrophysical constraints on the Hu-Sawicki gravity scenario in the metric formalism have been examined in a number of papers (see, e.g., Refs.~\cite{HS-papers}). However, there is a complete lack of studies on the mechanism of this proposal in the Palatini approach. Note that, irrespective of the formalism adopted, the model (\ref{Hu-Sawicki}) must satisfy certain viability conditions, as for example, the positivity of the effective gravitational coupling which requires $ \kappa^2 / f_R > 0 $ (to avoid anti-gravity). For the current epoch, this condition  implies
\begin{equation}
\label{c1c3}
f_{R0} = \frac{3 + 2(\frac{R_0}{m^2})^2c_2}{\frac{R_0}{m^2} \left[1 + 2(\frac{R_0}{m^2})c_2\right]} > 0\;,
\end{equation}
which corresponds to the following bounds for the parameter $c_2$:
\begin{equation}
\label{c1c2}
c_2 < -\frac{1}{2}\,\frac{m^2}{R_0} \quad \mbox{or}\quad c_2 > - \frac{3}{2} \left( \frac{ m^2 }{ R_0 } \right)^2\;.
\end{equation}
\section{Observational constraints}

In order to perform the observational tests discussed below, we solve Eq.~(\ref{eqf}) by taking the following steps: i) we first set $z = 0$ and compute $c_1$ (in terms of $c_2$ and $R_0$) from Eq.~(\ref{eq:trace}); ii) we combine the latter result with Eq.~(\ref{eqf}) in order to obtain a solution for $R_0$; iii) the value of $R_0$ is then used as initial condition to solve Eq.~(\ref{eq:trace}) by using a fourth-order Runge-Kutta method and, finally, to obtain the function $H(z)$, as given by (\ref{eqf}).

Fig.~\ref{fig:data_h-sn} shows the evolution of the Hubble parameter and the predicted distance modulus
$ \mu(z) = 5\log[d_L(z)/\text{Mpc}]+ 25 $, where
$ d_L = (1+z) \int_0^z \frac{ dz^{\prime} }{ H(z^{\prime}) }$ stands for the
luminosity distance, as a function of redshift for some best-fit values for
$\Omega_{m0}$ and $c_2$ obtained in our analyses (see Sec. IIIC). For the sake of comparison, the
standard $\Lambda$CDM prediction with $\Omega_{m0} = 0.26$ is also shown (dashed
line).

\subsection{Expansion rate}


 In order to test the observational viability of the
$f(R)$ scenario described by Eq. (\ref{Hu-Sawicki}), we firstly use current
measurements of the expansion rate at $ z \neq 0 $. Currently, this test is
based on the fact that Luminous Red Galaxies (LRGs) can provide us with direct
measurements of $H(z)$~\cite{Jimenez} (see also \cite{Ma} for a recent review on
$H(z)$ measurements from different techniques). This can be done by calculating
the derivative of cosmic time with respect to redshift, i.e.,
$ H(z) \simeq - \frac{1}{(1+z)} \frac{\Delta z}{\Delta t} $. This method was
first presented in~\cite{Jimenez} and consists in measuring the age difference
between two red galaxies at different redshifts in order to obtain the rate
$ \Delta z / \Delta t $. By using a recent released sample from Gemini Deep
Survey (GDDS)~\cite{gemini} and archival data~\cite{archival}, Ref.~\cite{newh}
has calculated 11 $H(z)$ data points over the redshift range
$ 0.1 \leq z \leq 1.75 $.

We estimate the free parameters $\mathbf{P}$ by using a $\chi^2$ statistics,
i.e.,
\begin{equation}
	\label{eq:chi2.h}
	\chi_H^2 = \sum_{i=1}^{N_H} \frac{\left[ H_{th}^i (z_i|\mathbf{P})	- H_{obs}^i (z_i) \right]^2}{\sigma^2_i} \,,
\end{equation}
where $H_{th}^i(z|\mathbf{P})$ is the theoretical Hubble parameter at redshift
$z_i$ and $\sigma_i$ is the uncertainty for each of the $N_H = 11$
determinations of $H(z)$ given in~\cite{newh}. In our analyses, the Hubble
constant $H_0$ is considered as a nuisance parameter and we marginalize over it.

\begin{figure*}[t]
	\includegraphics[width= 7.5cm, height=6.5cm]{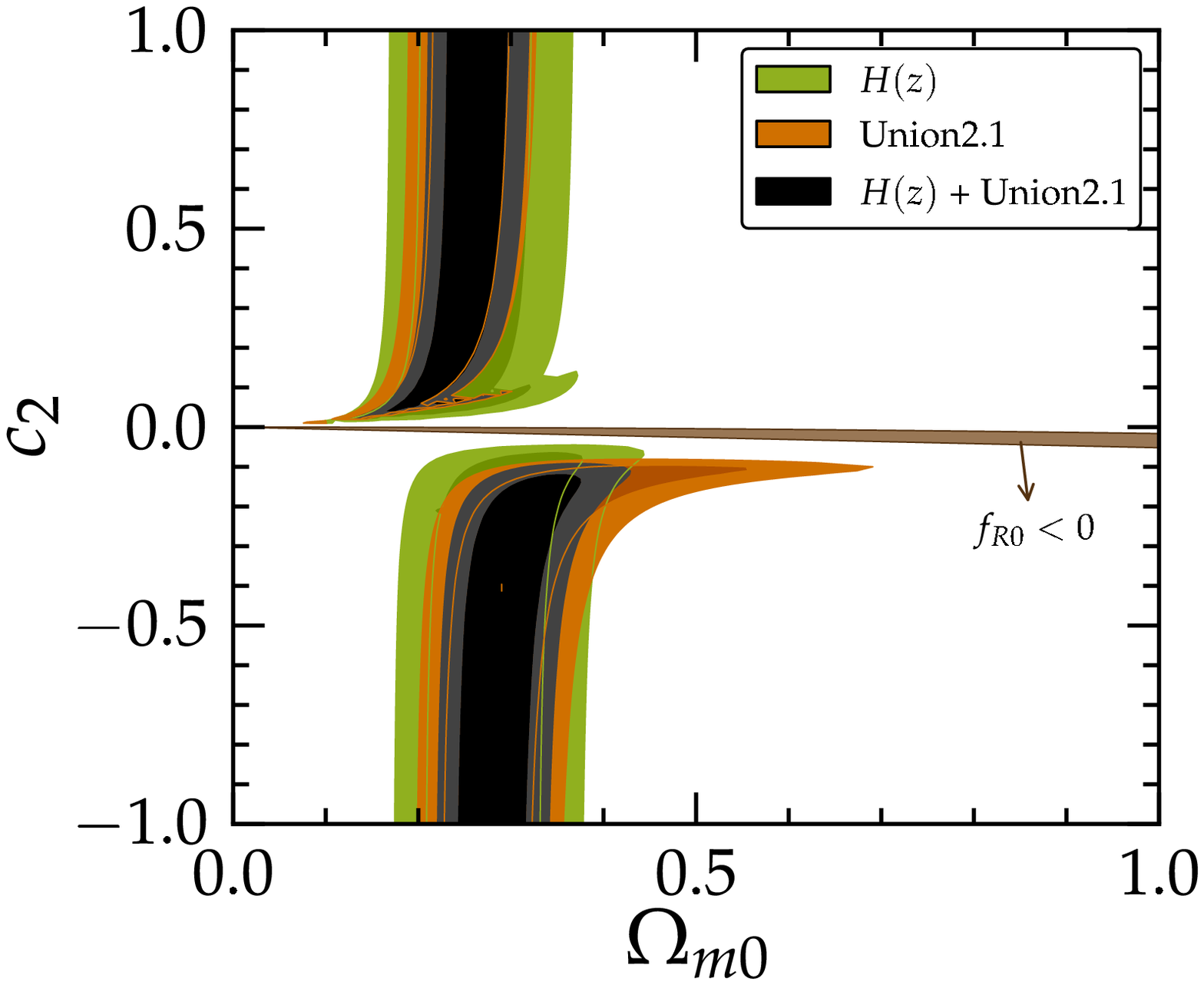}
	\includegraphics[width= 7.5cm, height=6.5cm]{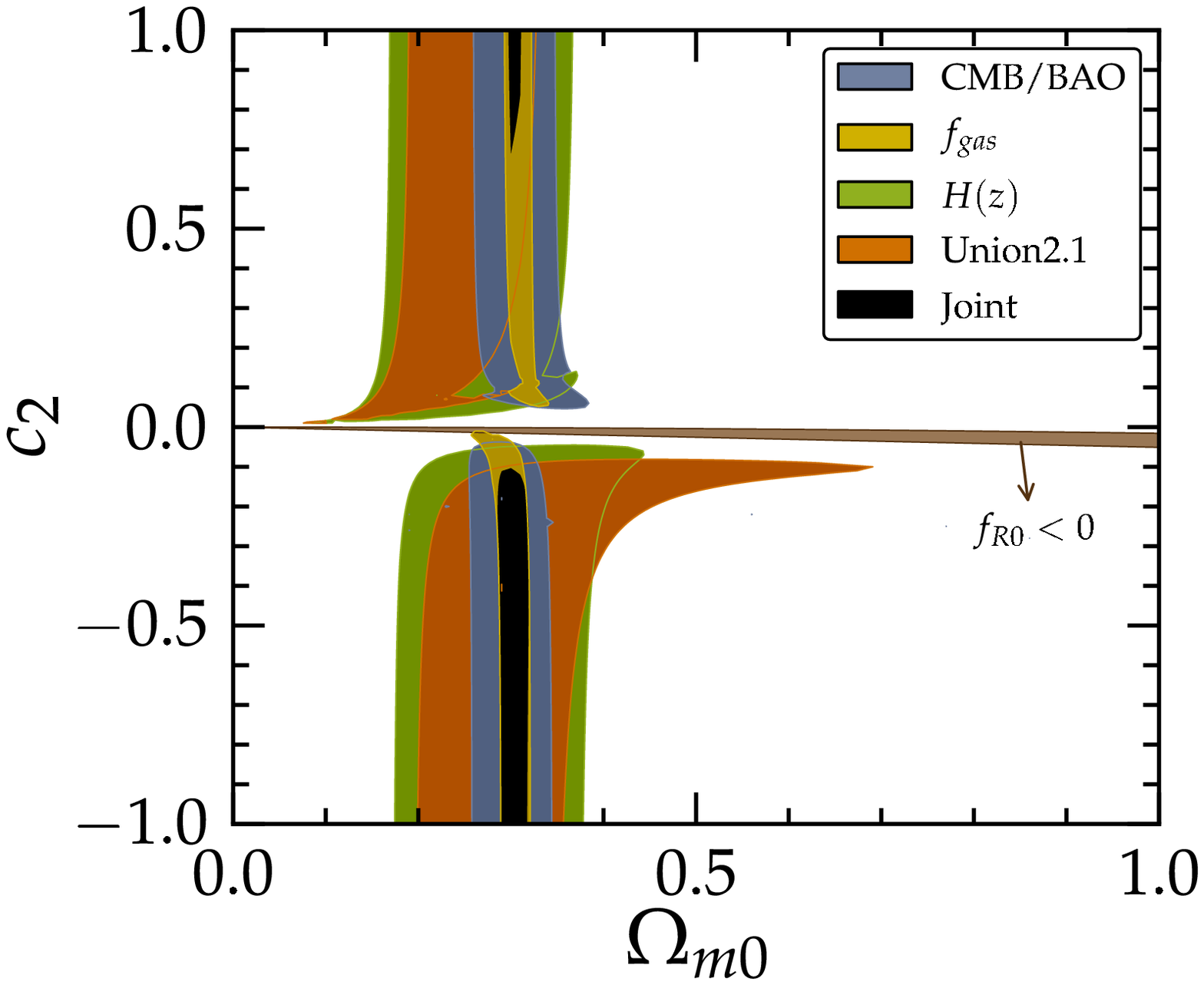}
	\caption{\label{fig:contours}
		(\textit{left}) Contours of $ \Delta \chi^2 = 2.3 $ and $ \Delta \chi^2 = 6.17 $ in the $ \Omega_{m0} \times c_2 $ plane arising from SNe Ia {union2.1} sample~\cite{union2.1} (Orange)
		and current $H(z)$ data~\cite{newh} (Green). Black contours
		stand for the joint analysis involving these two sets of data.
		(\textit{right})~Contours of $ \Delta \chi^2 = 6.17 $ in the
		$ \Omega_{m0} \times c_2 $ plane when 7 measurements of the
		CMB/BAO ratio and 57 measurements of the gas mass fraction of
		galaxy clusters are added. We have also indicated the region $f_{R0} < 0$ corresponding to the bounds (\ref{c1c2}) for the best-fit values shown in Table I.
	}
\end{figure*}

\begin{table*}
	\caption{\label{tab:results}
		Best-fit values for $c_2$, $c_1$ and $f_{R0}$. The error bars for $\Omega_{m0}$ correspond to 68.3\% and 95.4\% confidence intervals.}
	\begin{ruledtabular}
		\begin{tabular}{lllllll}
			Test					& $\Omega_{m0}$					& $c_2$		& $c_1$		& $f_{R0}$	& $\chi^2_{min}$	& $\chi^2_{r}$ \\
			\hline
			$H(z)$					& $0.22\;_{-0.08}^{+0.11}\;_{-0.13}^{+0.19}$	& 0.08		& 2.23			& 0.88		& 7.72			& 0.86 \\
			SNe Ia					& $0.29\;_{-0.17}^{+0.19}\;_{-0.21}^{+0.34}$	& -0.40		& -5.41			& 1.03		& 553.81		& 0.96 \\
			CMB/BAO					& $0.29\;_{-0.01}^{+0.03}\;_{-0.03}^{+0.07}$	& -0.18		& -2.17		& 1.08		& 1.24			& 0.25 \\
			$f_{gas}$				& $0.28\;_{-0.01}^{+0.01}\;_{-0.02}^{+0.04}$	& -0.03		& -0.04			& 1.61		& 95.10			& 1.73 \\
			$H(z)$ + SNe Ia				& $0.28\;_{-0.07}^{+0.06}\;_{-0.15}^{+0.12}$	& -0.85		& -12.64		& 1.02		& 561.63		& 0.95 \\
			CMB/BAO + $f_{gas}$ + $H(z)$ + SNe Ia	& $0.30\;_{-0.01}^{+0.01}\;_{-0.02}^{+0.02}$	& -0.25		& -3.04		& 1.06		& 659.87		& 1.01 \\
		\end{tabular}
	\end{ruledtabular}
\end{table*}

\begin{figure*}[t]
	\includegraphics[width= 7.5cm, height=6.0cm]{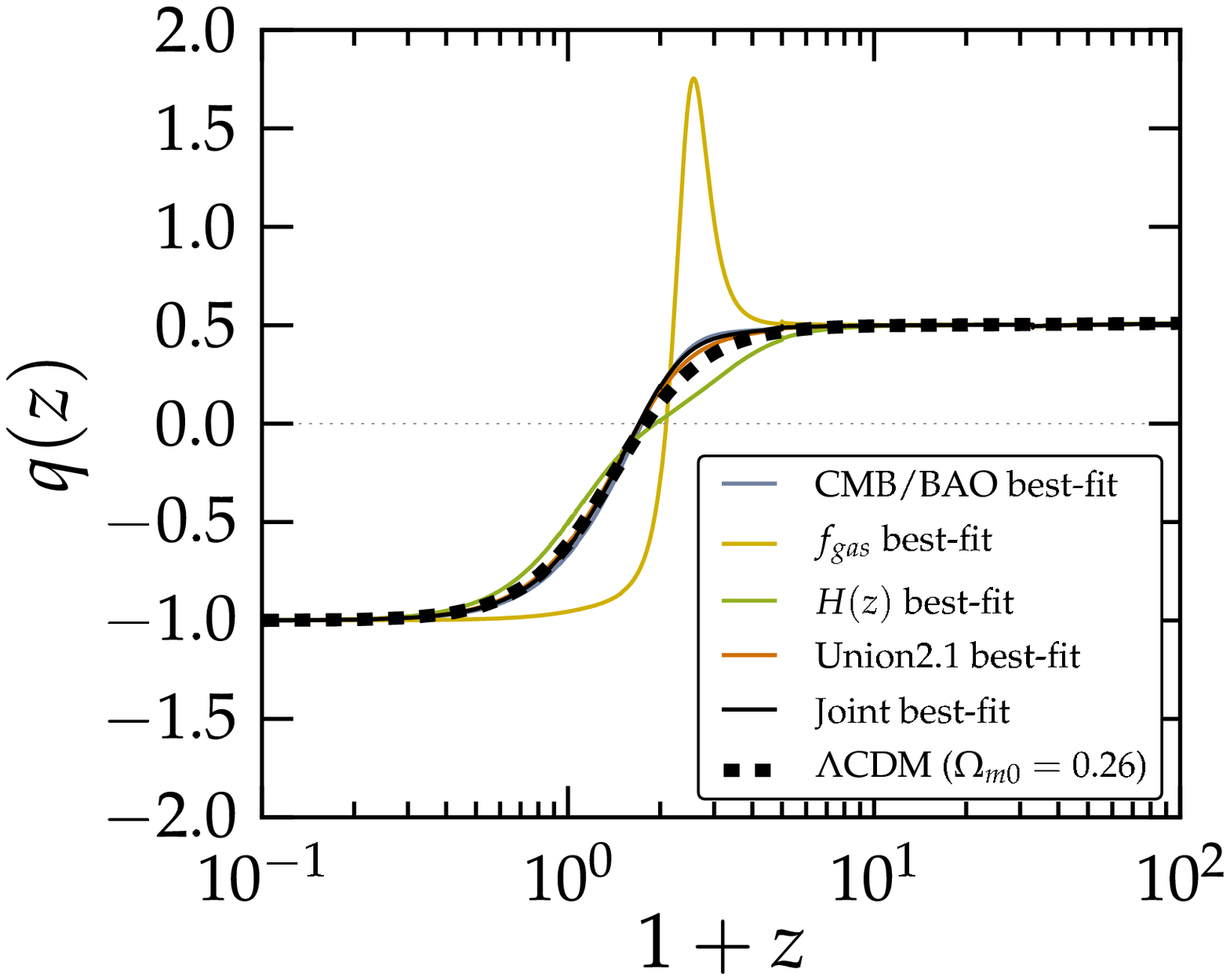}
	\includegraphics[width= 7.5cm, height=6.0cm]{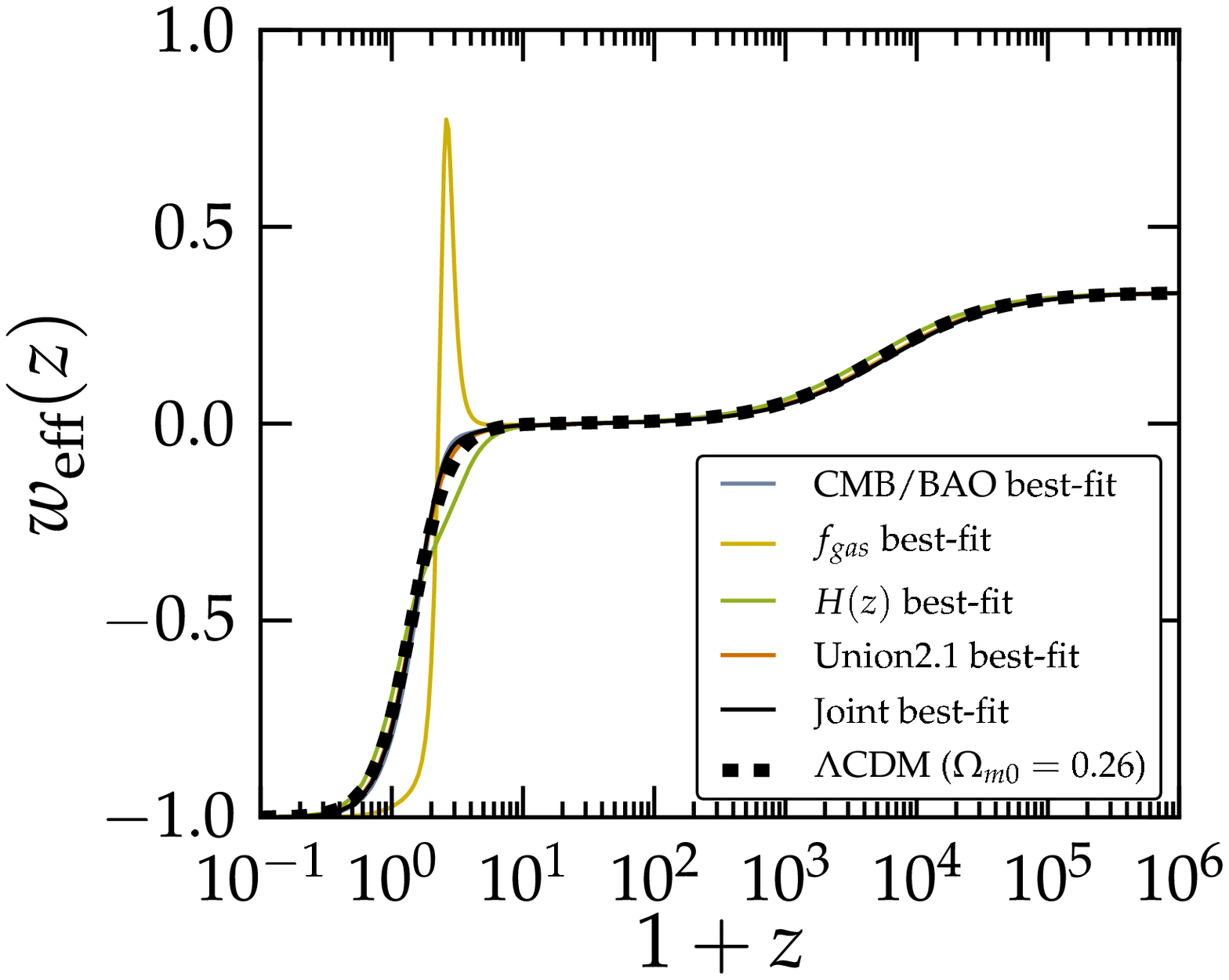}
	\caption{\label{fig:q_weff}
		 Deceleration parameter (\textit{left}) and effective equation of state  (\textit{right}) as a function of $z$ for the best-fit values of $\Omega_{m0}$ and $c_2$ presented in Table~\ref{tab:results}. The $\Lambda$CDM model is also shown for the sake of comparison.
	}
\end{figure*}

\subsection{SNe Ia}

 The predicted distance modulus for a supernova at redshift
$z$, given a set of parameters $\mathbf{P}$, is
$ \mu^p_i(z_i|\mathbf{P}) = 5\log{D_L(z_i,\mathbf{P})} + \mu_0 $, where
$ D_L = H_0 d_L $ is the Hubble-free luminosity distance,
$ \mu_0 = 42.38 - 5 \log_{10} h $ and $h$ stands for the Hubble constant $H_0$
in units of 100~km/s/Mpc. In our analysis, we  use the latest Union2.1 sample
which includes 580 SNe Ia over the redshift range $ 0.1 \leq z \leq 1.75 $. In
order to include the effect of systematic errors into our analyses, we
follow~\cite{union2.1} and estimated the best fit and error bars to the set of
parameters ${\mathbf{P}}$ by using a $\chi^2$ statistics, with
\begin{equation}
	\label{eq:chi2.sn}
	\chi^2_{SNe} = \sum_{i=1}^{580} ( \mu_i^{obs} - \mu_i^{p} ) ( {\mathbf{C}}^{-1} )_{ij} ( \mu_j^{obs} - \mu_j^{p} ) \,,
\end{equation}
where $\mathbf{C}$ is a 580 $\times$ 580 covariance matrix~\cite{union2.1}. As
for the analysis involving $H(z)$ data, we also marginalize over $H_0$.

\subsection{Results}

In Fig.~\ref{fig:contours}~(left) we show the first results of our statistical analyses. Contours corresponding to $ \Delta \chi^2 = 2.3 $ and $ \Delta \chi^2 = 6.17 $ in the $ \Omega_{m0} \times c_2 $ plane are shown for the $\chi^2$ given above. From this analysis, we clearly see that neither
the current $H(z)$ and SNe Ia measurements alone nor the combination of them ($ \chi^2  = \chi^2_{SNe} + \chi_H^2 $) can place tight constraints on the values of the $f(R)$ parameter $c_2$. We also observed that other cosmological observables seem to be unable to provide orthogonal contours to those shown in Fig.~\ref{fig:contours}~(left), which in principle would lead to tighter bounds on $c_2$ from a joint analysis. Fig.~\ref{fig:contours}~(right) shows the results obtained when two other sets of cosmological observations are included, namely: 7 estimates of the ratio of the CMB acoustic scale $\ell_A$ and the baryonic acoustic oscillation (BAO) peak, as given in Ref.~\cite{cmbbao}, and
the 57 measurements of the gas mass fraction in X-ray luminous galaxy clusters discussed in Ref.~\cite{fgas} (we refer the reader to Refs.~\cite{xray}
for more details on these cosmologial tests). For this joint analysis, the best-fit values are $ \Omega_{m0} = 0.30 $ and $ c_2 = -0.25 $, with the reduced $ \chi^2_r \equiv \chi^2_{min} / \nu \simeq 1.01 $ ($\nu$ is defined as degrees of freedom), which is very close to the value found for the standard $\Lambda$CDM model, $ \chi^2_r \simeq 1.001 $. This result clearly shows that these scenarios are indistinguishable from each other at the background level, which coincides  with the analysis performed in the metric formalism and discussed in Ref.~\cite{melchiorri} (although the expansion histories for both metric and Palatini formalisms are completely different). It is worth mentioning that in our analysis we consider both positive and negative values of the $f(R)$ parameter $c_2$, in agreement with the constraint (\ref{c1c2}).
Note also that for $c_2 = 0$, the cosmological scenario discussed above reduces to the Einstein-de Sitter model, being ruled out by current data. We use the joint best-fit value discussed above, as well as the others shown in Table~\ref{tab:results}, to discuss some cosmological consequences of this class of $f(R)$ cosmologies in the next section.

\section{Cosmological consequences}

 Two important quantities directly related
to the expansion rate of the Universe and its derivatives are the deceleration
parameter
\begin{subequations}
	\begin{equation}
		\label{q-z}
		q(z) = \frac{(1+ z)}{H(z)} H'(z) - 1 \;,
	\end{equation}
	and the effective equation of state (EoS)
	\begin{equation}
		w_{eff} = -1 + \frac{2(1+z)}{3H} H'(z) \;,
	\end{equation}
\end{subequations}
where a prime denotes differentiation with respect to $z$ and $H(z)$ is given by
Eq. (\ref{eqf}). For all sets of best-fit values displayed in
Table~\ref{tab:results}, we show $q(z)$ and $w_{eff}(z)$ as a function of the
redshift in Fig.~\ref{fig:q_weff}. Similarly to the analyses
presented in Sec. III, we have included a component of radiation
($\Omega_{r0} = 5 \times 10^{-5}$) to plot these curves. As can be seen from
Fig.~\ref{fig:q_weff}~(left), for some combinations of parameters the cosmic
evolution is well-behaved, with a past cosmic deceleration and a current
accelerating phase starting at $ z \simeq 1 $, which seems to be in good agreement with current kinematic analyses (see, e.g., \cite{kinematic}).

For all best-fit combinations shown in Table~\ref{tab:results},
Fig.~\ref{fig:q_weff}~(right) shows that the Universe went through the last
three phases of cosmological evolution, i.e., a radiation-dominated
($ w_{eff} = 1/3 $), a matter-dominated ($ w_{eff} = 0 $) and a late-time
accelerating phase ($ w_{eff} < 0 $), which is similar to what happens in the
standard $\Lambda$CDM cosmology.  From these results, it is also clear that the
arguments of Ref.~\cite{Amendola} about the behavior of $w_{eff}$ in the metric
approach (we refer the reader to \cite{cap} for a different conclusion) seems
not to apply to the Palatini formalism, at least for the class of models
discussed in this paper and the interval of parameters $\Omega_{m0}$ and $c_2$
given by our statistical analysis -- the same also happens to power-law~\cite{Tavakol,janilo} and
exponential~\cite{expogravity} $f(R)$ gravities in the Palatini formalism. Note also
that, differently from exponential-type $f(R)$ models (see, e.g.,
\cite{expogravity}), we have not found solutions of transient cosmic acceleration in
which the large-scale modification of gravity will drive the Universe to a new
matter-dominated era in the future.
\\

\section{Conclusions}

$f(R)$-gravity provides an alternative way to explain the current cosmic acceleration with no need of invoking either the existence of an extra spatial dimension or an exotic component of dark energy. Among a number of $f(R)$ models discussed in the literature,  the so-called Hu-Sawicki scenarios are designed to posses a chameleon mechanism that allows to evade solar system constraints. Although the cosmological scenario that arises from the metric formalism of this model has been shown to satisfy the conditions needed to produce a cosmologically viable expansion history, its Palatini version had not yet been  investigated.

In this paper, we have discussed several cosmological consequences of this class of models in the Palatini formalism. We have performed consistency checks and tested the observational viability of these scenarios by using one of the latest SNe Ia data, the so-called Union2.1 sample, measurements of the expansion rate $H(z)$ at intermediary and high-$z$, estimates of the CMB/BAO ratio and current observations of the gas mass fraction in X-ray luminous galaxy clusters. We have found a good agreement between these observations and the theoretical predictions of the model, with the reduced $ \chi^2_{min} / \nu \simeq 1 $ for the  analyses performed (see Table I). In what concerns the past cosmic evolution predicted by this class of models, we have shown that for a subsample of model parameters, the so-called Hu-Sawicki scenarios in the Palatini approach are indistinguishable from the standard $\Lambda$CDM model, being able to produce the sequence of radiation-dominated, matter-dominated, and accelerating periods without need of dark energy.

\textit{Acknowledgments} -- The authors thank CNPq, CAPES and FAPERJ for the grants under which this work was carried out. J.S. and J.S.A. also thank financial support from INCT-INEspa\c{c}o.


\end{document}